\documentclass[aps,prx,
 amsmath, twocolumn, amssymb,superscriptaddress,
 aps, physrev,
]{revtex4-2}

\usepackage{graphicx}
\usepackage{dcolumn}
\usepackage{bm}
\usepackage{tikz}
\usepackage{quantikz}
\usepackage{physics}
\usepackage{hyperref}
\usepackage[capitalize]{cleveref}
\usepackage{siunitx}

\usepackage{todonotes}

\crefname{equation}{Eq.}{Eqs.}
\Crefname{equation}{Eq.}{Eqs.}

\crefname{figure}{Fig.}{Figs.}
\Crefname{figure}{Fig.}{Figs.}

\usepackage{placeins}
\usepackage{mathtools}

\usepackage{csquotes}
\usepackage[normalem]{ulem}

\usepackage{orcidlink} 

\begin{document}

\title{Sample-Based Krylov Quantum Diagonalization for the Schwinger Model on Trapped-Ion and Superconducting Quantum Processors}

\author{Emil Otis Rosanowski~\orcidlink{0009-0002-4072-2801}}
\email{rosanowski@hiskp.uni-bonn.de}
\affiliation{
 Transdisciplinary Research Area “Building Blocks of Matter and Fundamental Interactions” (TRA Matter) and Helmholtz Institute for Radiation and Nuclear Physics (HISKP), University of Bonn, Nussallee 14-16, 53115 Bonn, Germany
}
\author{Jurek Eisinger~\orcidlink{0009-0008-9144-9052}}
\email{jeisinge@uni-mainz.de}
\affiliation{
 QUANTUM, University of Mainz, Department of Physics, Staudingerweg 7, Germany
}

\author{Lena Funcke~\orcidlink{0000-0001-5022-9506}}

\affiliation{
 Transdisciplinary Research Area “Building Blocks of Matter and Fundamental Interactions” (TRA Matter) and Helmholtz Institute for Radiation and Nuclear Physics (HISKP), University of Bonn, Nussallee 14-16, 53115 Bonn, Germany
}
\author{Ulrich Poschinger~\orcidlink{0000-0001-5341-7860}}
\affiliation{
 QUANTUM, University of Mainz, Department of Physics, Staudingerweg 7, Germany
}
\author{Ferdinand Schmidt-Kaler~\orcidlink{0000-0002-5697-2568}}
\affiliation{
 QUANTUM, University of Mainz, Department of Physics, Staudingerweg 7, Germany
}

\date{\today}

\begin{abstract}
We apply the recently proposed Sample-based Krylov Quantum Diagonalization (SKQD) method to lattice gauge theories, using the Schwinger model with a $\theta$-term as a benchmark. SKQD approximates the ground state of a Hamiltonian, employing a hybrid quantum–classical approach: (i)~constructing a Krylov space from bitstrings sampled from time-evolved quantum states, and (ii)~classically diagonalizing the Hamiltonian within this subspace. We study the dependence of the ground-state energy and particle number on the value of the $\theta$-term, accurately capturing the model’s phase structure. The algorithm is implemented on trapped-ion and superconducting quantum processors for system sizes ranging from $N=4$ to $N=30$ qubits, demonstrating consistent performance across platforms. We show that SKQD reduces the effective Hilbert space by up to a factor of \num{6.8 e-6}, and although the Krylov space dimension still scales exponentially, its slower growth underscores the method's potential for simulating lattice gauge theories in larger volumes.
\end{abstract}

\maketitle

\section{Introduction}
\label{sec:introduction}

Efficient and accurate determination of the low-energy spectrum of a Hamiltonian is a central challenge in particle and condensed-matter physics, quantum chemistry, and materials science, and serves as a key benchmark for quantum simulation and algorithm development. Exact diagonalization of the Hamiltonian becomes exponentially costly with system size on classical computers. The best-known quantum algorithm, quantum phase estimation~\cite{Kitaev:1995qy}, remains impractical on current noisy intermediate-scale quantum (NISQ) devices~\cite{Preskill:2018jim} due to the large circuit depths required (see, e.g., Refs.~\cite{Lin:2021rwb, Ding:2022xue}). Hybrid quantum-classical methods, such as the Variational Quantum Eigensolver (VQE)~\cite{Peruzzo:2013bzg}, offer greater hardware efficiency but rely on noisy parametric optimization and extensive sampling of observables, resulting in unfavorable scaling, high runtime, and Barren plateaus~\cite{Wang:2020yjh, Anschuetz:2022wvo}.

As an alternative approach, Krylov Quantum Diagonalization (KQD) has been proposed for probing spectral properties on pre-fault-tolerant quantum devices~\cite{Parrish:2019ruc,Motta:2019yya,Evangelista:2019ztq,deJong:2020oid,Cohn:2021pls,Seki:2021oxk,Baker:2021uul,Baker:2021ylc,Klymko:2021xoq,Jamet:2022ppb,Lee:2023xzf,Kirby:2022ncy,Shen:2022lmk,Tkachenko:2022vwn,Epperly:2021ugt,Kirby:2024roq}. In KQD, a Krylov space is built by evolving a reference state over a series of time intervals, after which the Hamiltonian is diagonalized classically within this subspace. The method uses time-evolution circuits to reach system sizes inaccessible to classical simulation, while ensuring convergence when the initial state has a polynomial overlap with the true ground state, meaning that the initial state's overlap with the ground state does not shrink exponentially with system size but instead decreases only as an inverse polynomial~\cite{Kim:2023bwr,Shinjo:2024vci}. KQD has been used to compute eigenenergies of quantum many-body systems on lattices of up to 56 sites~\cite{Yoshioka:2024lle} and to extract ground-state properties of the lattice Schwinger model~\cite{Anderson:2024kfj}. 

Building on these ideas, the \textit{Sample-Based} Krylov Quantum Diagonalization (SKQD) method has been recently introduced~\cite{Yu:2025czp}. SKQD combines KQD with key concepts from subspace algorithms using independent samples from a quantum computer~\cite{Kanno:2023rfr,Robledo-Moreno:2024pzz,Kaliakin:2024edc}: estimating ground-state energies by sampling quantum states and performing classical post-processing and diagonalization on noisy data. Specifically, SKQD constructs a subspace from bitstrings sampled from time-evolved reference states and diagonalizes the Hamiltonian classically within this subspace to approximate the ground state. This hybrid approach inherits the noise resilience of sample-based methods while preserving the convergence guarantees of Krylov methods. SKQD has been proven to approximate the ground-state energy in polynomial time, assuming sparsity of the ground state and that the reference state has polynomial overlap with the true ground state~\cite{Yu:2025czp}. The method has been demonstrated experimentally for quantum many-body systems with up to 41 sites~\cite{Yu:2025czp}.

In this paper, we present the first application of SKQD to lattice gauge theories, in particular, to the Schwinger model~\cite{Schwinger:1962tp}. 
The Schwinger model describes quantum electrodynamics in 1+1 dimensions and exhibits QCD-like features such as confinement, chiral symmetry breaking, and a topologically non-trivial vacuum with a $\theta$-term. Owing to these properties, its lattice formulation serves as a common benchmark for developing and testing numerical methods aimed at Lattice QCD applications, including new approaches to address the sign problem, tensor network techniques, and quantum algorithms (see, e.g., Refs.~\cite{Banuls:2018jag,Funcke:2023jbq} for reviews). 

Specifically, we investigate the dependence of the ground-state energy and particle number on the $\theta$-term, aiming to explore the first-order phase transition that occurs in the Schwinger model at $\theta = \pi$~\cite{Coleman:1976uz,Hamer:1982mx, Buyens:2017crb, Byrnes:2002nv}. For the quantum hardware implementations, we execute SKQD on both a trapped-ion quantum processor~\cite{Kaushal:2019skm} and the IBM superconducting processors \texttt{ibm\_marrakesh} and \texttt{ibm\_kingston}~\cite{ibm_quantum_processor_types}. Trapped-ion devices offer enhanced coherence times and full qubit connectivity, whereas superconducting processors currently provide faster gate times and better scalability. Accordingly, in this study, the trapped-ion processor serves as a proof-of-principle demonstration, using $N=4$ qubits, while the IBM devices illustrate the scaling of the method to larger systems with up to $N=30$ qubits.

We observe excellent agreement with exact diagonalization results, even when using only the kinetic part of the Schwinger Hamiltonian for the time evolution in the SKQD algorithm. This simplified time evolution significantly improves the tractability of the problem on currently available NISQ platforms. Notably, although the Krylov space has a much smaller dimension than the full Hilbert space, it still allows for accurate reconstruction of energies and phase transitions. SKQD reduces the effective Hilbert space by up to a factor of \num{6.8 e-6} while keeping the relative energy deviations to the true ground state in the order of \num{e-3}. The Krylov space dimension still scales exponentially, but its slower growth highlights the potential of this approach for simulating larger quantum systems.

The paper is organized as follows: In \cref{sec:schwinger}, we introduce the Schwinger model. The SKQD method is then discussed in \cref{subsec:sample-based-krylov}, followed by details on its application to the Schwinger model in \cref{subsed:TEwKH}. Results from the trapped-ion quantum computer are presented in \cref{subsec:tiqc}, while results from the IBM devices are discussed in \cref{subsec:IBM}. Finally, in \cref{sec:PTlarge}, we investigate the phase transition using SKQD with a specific ground-state initialization of the mass Hamiltonian for systems up to $N=30$ qubits, where exact diagonalization is not easily available. We conclude in \cref{sec:conclusion}. In Appendix~\ref{sec:QC}, we provide a detailed description of the trapped-ion quantum processor~\cite{Kaushal:2019skm} used in this work. Additional results for the ground-state energy at different qubit numbers, completing the scaling analysis in \cref{subsec:IBM}, are presented in Appendix~\ref{app:16_18}. In Appendix~\ref{app:SubspaceDim}, we show further plots on the scaling of the Krylov space dimension with the number of Trotter steps.

\section{The Schwinger Model}
\label{sec:schwinger}

\subsection{Continuum Formulation}
\label{sec:cont}

In the continuum, the Hamiltonian density of the Schwinger model~\cite{Schwinger:1962tp} with a single fermion flavor and a $\theta$-term is given by
\begin{equation}
\label{hamiltonian_density}
    \mathcal{H} = -i\overline{\psi}\gamma^1\left(\partial_1 - igA_1\right)\psi + m\overline{\psi}\psi + \frac{1}{2}\left(\dot{A}_1 + \frac{g\theta}{2\pi}\right)^2\,,
\end{equation}
where $\psi$ is a two-component spinor, $m$ the bare fermion mass, $g$ its coupling to the gauge field, and $\theta$ the parameter of the topological $\theta$-term, corresponding to a constant background electric field~\cite{Coleman:1976uz}. Temporal gauge ($A_0=0$) has been imposed, so that only the $F_{01}=\dot{A_1}$ component of the field strength appears. The physically relevant states have to fulfill Gauss's law
\begin{align}
    -\partial_1\dot{A}^1 = g\overline{\psi}\gamma^0\psi\,,
    \label{eq:gauss_law}
\end{align}
which relates the spatial variation of the electric field to the local charge density.

The $\theta$-term is $2\pi$-periodic and above a certain critical mass $m_c/g$, a first-order phase transition can be observed \cite{Coleman:1976uz}. The transition becomes second-order at exactly this mass, which was numerically found to be $m_c/g\approx0.33$ \cite{Hamer:1997dx,Buyens:2017crb,Byrnes:2002nv}.

In the presence of a topological $\theta$-term, a first-order phase transition occurs in the Schwinger model at $\theta = \pi$ for fermion masses above a critical value $m_c/g$~\cite{Coleman:1976uz}, which was numerically determined to be $m_c/g\approx0.33$ \cite{Hamer:1982mx, Buyens:2017crb, Byrnes:2002nv}. For large $m/g \gg 1$, the creation of charged particles is suppressed. When 
$\theta < \pi$, the vacuum remains uncharged and the electric field is given by the background value. For 
$\theta > \pi$, it becomes energetically favorable to create a particle–antiparticle pair connected by a flux string, reducing the bulk electric field from $\theta/2\pi$ to $\theta/2\pi - 1$ according to Gauss’s law~\eqref{eq:gauss_law}. At $\theta =\pi$, these two states are degenerate, since the electric field energy scales with the square of the electric field and the particle mass becomes negligible in the infinite-volume limit. The resulting phase transition at $\theta=\pi$ is therefore accompanied by a change in particle number, and a spontaneous breaking of charge-conjugation–parity (CP) symmetry~\cite{Coleman:1976uz}.

\subsection{Lattice Discretization}
\label{subsec:LatticeDiscretization}
In this work, we focus on the staggered fermion discretization, and correspondingly consider the lattice version of the continuum Schwinger Hamiltonian in \Cref{hamiltonian_density} given by the Kogut–Susskind Hamiltonian~\cite{Kogut:1974ag}
\begin{align}
\label{Staggered_hamiltonian_initial}
    \begin{aligned}
        H =& -\frac{i}{2a}\sum_{n=0}^{N-2} \left(\phi^\dagger_n U_n\phi_{n+1}-\text{h.c}.\right)  \\
        &+ m_\text{lat}\sum_{n=0}^{N-1} (-1)^n\phi^\dagger_n\phi_n + \frac{ag^2}{2}\sum_{n=0}^{N-2} \left(L_n+l_0\right)^2 \,,
    \end{aligned}
\end{align}
where $\phi_n$ denotes a single-component fermionic field at site $n$, $m_{\rm lat}$ is the bare lattice fermion mass, $g$ the coupling constant, $a$ the lattice spacing, and $N$ the (even) number of lattice sites. The operators $U_n$ and $L_n$ act on the gauge links between fermionic sites, with $U_n$ serving as a raising operator for the quantized electric flux $L_n$ on link $n$. The constant background electric field is given by $l_0 = \theta / 2\pi$.

On the lattice, the Gauss's law constraint on the physical states reads
\begin{equation}
    L_n - L_{n-1} = Q_n \forall n\,,
    \label{eq:latGauss}
\end{equation}
where $Q_n = \phi^\dagger_n\phi_n-(1-(-1)^n)/2$ is the staggered fermionic charge operator.
For open boundary conditions, the electric field can be integrated out using Gauss's law \eqref{eq:latGauss}, and a residual gauge transformation can be applied, yielding a purely fermionic lattice Hamiltonian~\cite{Hamer:1997dx,Banuls:2013jaa}
\begin{align}
    \begin{aligned}
            H =& -\frac{i}{2a}\sum_{n=0}^{N-2} \left(\phi^\dagger_n \phi_{n+1}-\text{h.c}.\right)  \\
            &+ m_\text{lat}\sum_{n=0}^{N-1} (-1)^n\phi^\dagger_n\phi_n + \frac{ag^2}{2}\sum_{n=0}^{N-2}\left(l_0+ \sum_{k=0}^{n} Q_k\right)^2\,.
    \end{aligned}
    \label{eq:Staggered_hamiltonian_integrated}
\end{align}
Applying a Jordan–Wigner transformation~\cite{Jordan:1928wi}, the fermionic degrees of freedom can be mapped to spins, yielding the dimensionless lattice Hamiltonian~\cite{Hamer:1997dx,Banuls:2013jaa}
\begin{align}
    \label{Staggered_hamiltonian}
        \begin{aligned}
        W =&\; \frac{x}{2}\sum_{n=0}^{N-2} \left(X_n X_{n+1}+ Y_n Y_{n+1}\right) \\
        &+\frac{m_\text{lat}}{g}\sqrt{x}\sum_{n=0}^{N-1} (-1)^n Z_n
        + \sum_{n=0}^{N-2} \left(l_0 + \sum_{k=0}^{n}Q_k\right)^2 \,,
        \end{aligned}
\end{align}
where $x = 1/(a g)^2$ defines the inverse lattice spacing squared in units of the coupling, and $X_n, Y_n, Z_n$ are the Pauli operators acting on spin $n$. The staggered charge is now $Q_k=(Z_k+(-1)^k)/2$.
In the following, we will fix $N/\sqrt{x}=30$ in order to facilitate comparison with the results of \citet{Angelides:2023noe}.

In the continuum, only states with vanishing total charge have finite energy~\cite{Melnikov:2000cc}. To restrict our analysis to this physical sector, we add a penalty term to the Hamiltonian,
\begin{equation}
    \lambda\left(\sum_{n=0}^{N-1}Q_n\right)^2,
    \label{eq:penalty}
\end{equation}
where the coefficient $\lambda$ is chosen sufficiently large.

Expressing the penalty term in \Cref{eq:penalty} and the electric field energy in \Cref{Staggered_hamiltonian} in terms of Pauli matrices, with $Q_n=(Z_n+(-1)^n)/2$, the final Hamiltonian in terms of Pauli operators becomes \cite{Angelides:2023noe}
\begin{align}
\label{Staggered_hamiltonian_final}
    \begin{aligned}
        W =&\; \frac{x}{2}\sum_{n=0}^{N-2} \left(X_n X_{n+1}+ Y_n Y_{n+1}\right) \\
        &+\frac{m_\text{lat}}{g}\sqrt{x}\sum_{n=0}^{N-1} (-1)^n Z_n\\
        &+\frac{1}{2}\sum_{n=0}^{N-2}\sum_{k=n+1}^{N-1}(N-k-1+\lambda)Z_nZ_k\\
        &+\sum_{n=0}^{N-2}\left(\frac{N}{4}-\frac{1}{2}\left\lceil\frac{n}{2}\right\rceil+l_0(N-n-1)\right)Z_n \\
        &+l_0^2(N-1) + \frac{1}{2}l_0N + \frac{1}{8}N^2 + \frac{\lambda}{4}N,
    \end{aligned}
\end{align}
where $\lceil \cdot \rceil$ denotes the ceiling function. This Hamiltonian can be efficiently treated using the SKQD method, as discussed in \Cref{sec:methods}.

\subsection{Observables}
\label{subsec:Observables}

To identify the first-order phase transition in $\theta$ (or equivalently the constant background field $l_0$), we follow \citet{Angelides:2023noe} and use the particle number as an order parameter, since it displays a discontinuity at the transition. The particle number operator is defined as
\begin{equation}
\label{eq:particlenumber}
    P = \frac{N}{2}+ \sum_{n=0}^{N-1} (-1)^n\phi^\dagger_n\phi_n = \frac{N}{2}+ \sum_{n=0}^{N-1} (-1)^nZ_n\,,
\end{equation}
which is equivalent to the chiral fermion condensate up to a constant shift \cite{Angelides:2023noe}. In the absence of particles (i.e., spins down on even sites and spins up on odd sites), the sum in \Cref{eq:particlenumber} contributes $-1$ per site, whereas the opposite configuration contributes $+1$ per site. The first term in \Cref{eq:particlenumber} therefore ensures that $P$ is non-negative, so that the particle number is positive semidefinite.

In the continuum, the phase transition occurs at a critical value of $\theta_c=\pi$ (see \Cref{sec:cont}), corresponding to $l_{0,c} = 1/2$. On a finite lattice, however, this critical value is shifted to larger $l_{0,c}$. For small values of the background field $l_0$ and large fermion mass $m_\text{lat}/g$, particle creation is energetically suppressed, leading to $\langle P \rangle = 0$. As $l_0$ increases, it becomes energetically favorable to create a negatively charged particle on the left and a positively charged particle on the right boundary of the lattice, connected by a flux string~\cite{Coleman:1976uz, Byrnes:2002nv}. Consequently, the particle number exhibits a discontinuous jump to $\langle P \rangle = 2$ once $l_{0,c}$ is exceeded.

\section{Methods}
\label{sec:methods}

\subsection{Sample-Based Krylov Quantum Diagonalisation}
\label{subsec:sample-based-krylov}

We employ a sample-based variant of the Krylov Quantum Diagonalization (KQD) algorithm~\cite{Parrish:2019ruc, Motta:2019yya, Evangelista:2019ztq, deJong:2020oid, Cohn:2021pls, Seki:2021oxk, Baker:2021uul, Baker:2021ylc, Klymko:2021xoq, Jamet:2022ppb, Lee:2023xzf, Kirby:2022ncy, Shen:2022lmk, Tkachenko:2022vwn, Epperly:2021ugt, Kirby:2024roq}, referred to as Sample-Based Krylov Quantum Diagonalization (SKQD) \cite{Yu:2025czp}. In the original KQD framework, the central idea is to construct a Krylov subspace $\mathcal{K}$ and then diagonalize the Hamiltonian $H$ within that subspace. The Krylov subspace can be generated in various ways, such as the \textit{power Krylov space} (commonly used in classical numerical methods and also in quantum algorithms, see, e.g.,~\citet{Anderson:2024kfj}) or the \textit{unitary Krylov space} (particularly suited for simulating quantum time evolution, see, e.g.,~\citet{Yoshioka:2024lle}).

SKQD follows the same underlying idea as KQD but constructs the Krylov subspace in a different manner. Instead of generating basis vectors through repeated applications of time-evolution operators, SKQD samples a set of bitstrings  from time-evolved quantum states. These bitstrings are identified with computational basis states $\{\ket{b}\}$, which span the Krylov space as $\mathcal{K} = \mathrm{span}\big(\{\ket{b}\}\big)$. In this way, the sampled basis approximates the \textit{unitary Krylov space} discussed above. The Hamiltonian $H$ is then diagonalized within this subspace.

The SKQD protocol proceeds as follows~\cite{Yu:2025czp}:

\begin{enumerate}
\item On a quantum computer, prepare an initial reference state (e.g., $\ket{\psi_0} = \ket{0011}$ for four qubits, which is the initial state chosen in \cref{subsec:tiqc}).
\item Perform time evolution of the reference state:
\begin{equation}
    \label{meq:krylov-basis}
    \ket{\psi_k} = \left(e^{-i H\, dt}\right)^k \ket{\psi_0} \equiv \left[U(dt)\right]^k \ket{\psi_0},
\end{equation}
where $H$ is the Hamiltonian, $dt$ is the time step, $k \in \{0,1,\dots,K\}$ labels the evolution step, and $U(dt)$ is the unitary time-evolution operator.
\item For each step $k = 1, \dots, K$, execute the quantum circuit preparing $\ket{\psi_k}$ in steps 1 and 2 a sufficiently large number of times (called the number of shots $n_{\rm shots}$) and record the measurement outcome for each run. Upon discarding redundant bitstrings, $M$ uniquely sampled bitstrings remain. Obtain an approximate representation of $\ket{\psi_k}$ as a superposition of the $M$ uniquely sampled bitstrings:
\begin{equation}
    \ket{\psi_k} \approx \sum_{i=0}^{M-1} a_k^i \ket{b_k^i},
    \label{eq:psik}
\end{equation}
where $a_k^i$ is the unknown and irrelevant amplitude associated with the bitstring $\ket{b_k^i}$. This yields a set of uniquely sampled bitstrings at each step,
\begin{equation}
    \{\ket{b}\}_k = \{\ket{b_k^0}, \dots, \ket{b_k^{M-1}}\}.
\end{equation}
\item On a classical computer, construct the Krylov space by taking the span of all uniquely sampled bitstrings accumulated up to step $k$:
\begin{equation}
    \mathcal{K}_k = \mathrm{span}\big( \bigcup_{j \le k} \{\ket{b}\}_j \big),
\end{equation}
and project the Hamiltonian into this subspace:
\begin{equation}
    H^{\mathrm{P}}_{mn} = \bra{b_m} H \ket{b_n},
    \label{eq:HP}
\end{equation}
where $\ket{b_m}, \ket{b_n}\in \bigcup_{j \le k} \{\ket{b}\}_j$.
\item Finally, diagonalize the projected Hamiltonian $H^{\mathrm{P}}$ on a classical computer to obtain approximations of the energy eigenvalues and eigenstates.
\end{enumerate}

This procedure offers two key advantages. First, the algorithm is significantly more resilient to noise than alternative quantum algorithms for computing spectral properties of Hamiltonian, such as KQD, VQE, or quantum phase estimation. This robustness arises because only the basis vectors $\ket{b_k^i}$ of the Krylov space are required for the classical projection of the Hamiltonian (see \Cref{eq:HP}), and the exact amplitudes $a_k^i$ in \Cref{eq:psik} are irrelevant, provided that each bitstring $\ket{b_k^i}$ is sampled at least once. Consequently, the precise amplitudes of the computational basis states do not affect the projected Hamiltonian or the resulting eigenvalue accuracies. 

Second, the structure of the Schwinger model allows for classical postselection with respect to the zero-charge sector. Specifically, any measured bitstring corresponding to a nonzero total charge (e.g., $\ket{1011}$) can be discarded, ensuring that only physically valid states contribute to the final results.

\begin{figure}
    \centering
    \begin{quantikz}
        \lstick{$q_0$}& \gate[2]{R}\gategroup[4, steps = 3]{$U(dt)$} & & & \rstick[4]{$\ket{\psi_1}$}\\
        \lstick{$q_1$}& & \gate[2]{R} & &\\
        \lstick{$q_2$}&  & & \gate[2]{R} &\\
        \lstick{$q_3$}&  & & &
    \end{quantikz}
    \caption{Example circuit for preparing $\ket{\psi_k}$ in \Cref{eq:psik}, with $N=4$ qubits and a single Trotter step ($k=1$). Here, $U(dt)$ is the unitary time-evolution operator for a timestep $dt$, as defined in~\Cref{eq:Ugate}. The label $R$ denotes the $R_{XX+YY}$ gate as defined in Qiskit~\cite{Qiskit}.}
    \label{fig:ExampleCircuit}
\end{figure}
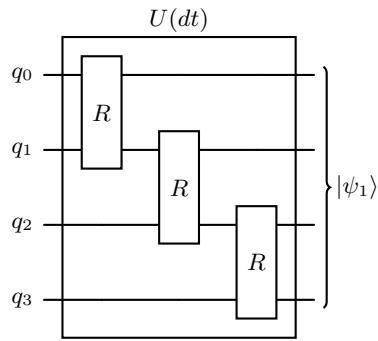

\subsection{Time Evolution with Kinetic Hamiltonian}
\label{subsed:TEwKH}

We now discuss the implementation of the unitary time-evolution operator $U(dt)$ in \Cref{meq:krylov-basis}. To minimize the circuit depth, we include only the kinetic part of the Hamiltonian in $U(dt)$, since the remaining terms are diagonal and do not generate superpositions of different states. This approximation allows us to construct a physically meaningful subspace using shallow quantum circuits. 
Empirically, we found that it is not necessary to implement the exact physical time evolution for the particular case of the Schwinger model. The goal is merely to generate basis states whose superpositions approximate the Hamiltonian’s ground state.

\begin{figure}
    \centering
    \includegraphics[width=\linewidth]{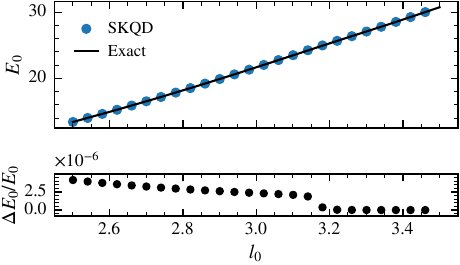}
    \caption{Ground-state energy $E_0$ of the Schwinger model as a function of the constant background field $l_0$, for $N=4$, \(N/\sqrt{x}=30\), and $m_{\rm lat}/g=10$. The upper panel shows SKQD results obtained on a trapped-ion quantum processor (blue) compared with exact diagonalization (black), while the lower panel displays their relative difference. Results correspond to a single Trotter step.}
    \label{fig:mainzQC_energies_4qb}
\end{figure}

Specifically, we define
\begin{equation}
    U(dt) = \exp{i\frac{dt}{4}\sum_{n=0}^{N-2}X_nX_{n+1}+Y_nY_{n+1}}\,,
\end{equation}
which we approximate by a first-order Trotter decomposition~\cite{Trotter1959,Suzuki:1976be} as
\begin{equation}
\label{eq:Ugate}
    U(dt)\approx \prod_{n=0}^{N-2}\exp{i\frac{dt}{4}\left(X_nX_{n+1}+Y_nY_{n+1}\right)}\,.
\end{equation}

Several such steps of the time evolution are performed, hereafter referred to as Trotter steps, corresponding to $\left[U(dt)\right]^k$ for the $k$-th step, see \Cref{meq:krylov-basis}. The corresponding quantum circuit is illustrated for four qubits and $k=1$ in \Cref{fig:ExampleCircuit}.

Since the number of Trotter steps required to reach a desired precision is not known a priori, we adopt a classical stopping criterion in our experiments (\Cref{sec:results}): during the diagonalization of the Hamiltonian in the subspace generated at each Trotter step, we stop once the relative improvement in energy over a large number of consecutive steps falls below a fixed threshold $c$. In this work, we choose $c=10^{-2}$ unless stated otherwise.

\section{Results using arbitrary State Initialization}
\label{sec:results}

We present SKQD results for the ground-state energy and particle number of the Schwinger model, focusing on the first-order phase transition discussed in \Cref{sec:schwinger}. Throughout, we choose a large fermion mass \(m_{\rm lat} /g= 10\), for which the phase transition is expected, and we scan over the background electric field \(l_0\). We furthermore keep the lattice volume fixed at \(N/\sqrt{x} = 30\) and vary the system size \(N\), corresponding to the number of qubits, between 4 and 20.

In \Cref{subsec:tiqc}, we show results obtained on a trapped-ion quantum processor~\cite{Kaushal:2019skm}, and in \Cref{subsec:IBM} results from the superconducting processors \texttt{ibm\_marrakesh} and \texttt{ibm\_kingston}~\cite{ibm_quantum_processor_types}. Trapped-ion devices offer excellent operational fidelities, while superconducting processors provide faster gate times and greater scalability. Accordingly, we use the trapped-ion platform for a proof-of-principle demonstration (\(N = 4\) qubits), and the superconducting devices to illustrate the scaling of SKQD to larger system sizes, with \(N = 14\)–\(20\) qubits. 

\subsection{Trapped-Ion Quantum Processor}
\label{subsec:tiqc}

\begin{figure}
    \centering
    \includegraphics[width=\linewidth]{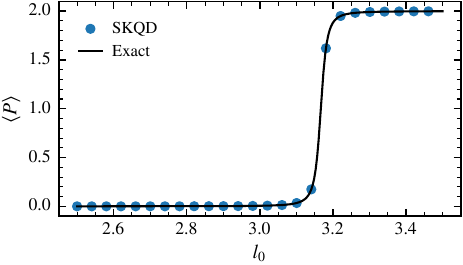}
    \caption{Particle number $\langle P\rangle$ of the Schwinger model as a function of the constant background field $l_0$, for $N=4$, \(N/\sqrt{x}=30\), and $m_{\rm lat}/g=10$. Shown are SKQD results obtained on a trapped-ion quantum processor (blue) compared with exact diagonalization (black). Results correspond to a single Trotter step.}
    \label{fig:mainzQC_ps_4qb}
\end{figure}

\begin{figure}
    \centering
    \includegraphics[width=\linewidth]{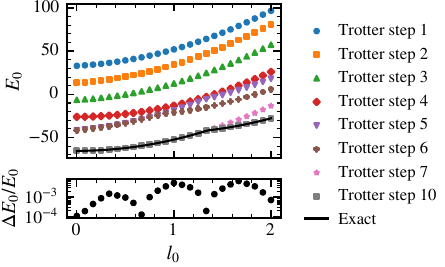}
    \caption{Ground-state energy \(E_0\) of the Schwinger model as a function of the constant background field \(l_0\), for \(N=14\), \(N/\sqrt{x}=30\), and \(m_{\rm lat}/g=10\). The upper panel shows SKQD results obtained on superconducting quantum processors for different numbers of Trotter steps (colored points) compared with exact diagonalization (black line). The lower panel displays the relative difference between the SKQD result at Trotter step 10 and the exact diagonalization result.}
    \label{fig:IBM_N_14_energy}
\end{figure}

As the first step in testing the SKQD algorithm for the Schwinger model, we implement it on a trapped-ion quantum processor~\cite{kaufmann2017scalable}. Experimental details are provided in Appendix~\ref{sec:QC}. The quantum circuit is initialized in the zero-charge state \(\ket{\psi_0}\equiv\ket{q_3 q_2 q_1 q_0} = \ket{0011}\). To select an appropriate time step \(dt\), we empirically chose \(dt = 0.77\). Other choices of $dt$ are possible; however, if $dt$ is chosen too small, successive steps yield nearly identical samples and the Krylov space grows only slowly. Conversely, if $dt$ is too large, Trotter errors accumulate and many sampled states become irrelevant, which degrades convergence. The evolution is carried out using a single Trotter step, and measurements are performed with \(n_\text{shots} = 400\) shots. 
The four-qubit quantum circuit is illustrated in \Cref{fig:ExampleCircuit}. 

The estimators for the probabilities $p_i$ of the outcomes $\ket{b_k^i}$ are the observed relative frequencies
\begin{equation}
    \Bar{p}_i = \frac{s_i + 1}{n_\text{shots} + 2^N}
\end{equation}
where $s_i$ are the observed counts for each event, subject to $\sum_i s_i = n_\text{shots}$. 
The associated confidence intervals are computed from the Laplace-corrected relative frequencies.

After the single Trotter step ($k=1$), the measured bitstrings \(\ket{b_k^i}\) are sampled with the following relative frequencies~\(\Bar{p}_i\):  
\begin{align}
\begin{split}
    &\ket{b_1^0} = \ket{0110}: \Bar{p}_0=\num{0.014}\pm\num{0.048}, \\ 
    &\ket{b_1^1} =  \ket{1001}: \Bar{p}_1=\num{0.029}\pm\num{0.050}, \\
    & \ket{b_1^2} =  \ket{0101}: \Bar{p}_2=\num{0.101}\pm\num{0.048}, \\ 
    &\ket{b_1^3} =  \ket{0011}: \Bar{p}_3=\num{0.409}\pm\num{0.055}.
\end{split}
\end{align}
Bitstrings for which the relative frequencies are statiscally consistent with zero are considered irrelevant and are therefore discarded.

This leaves the two dominant contributions
\begin{equation}
    \ket{b_1^2} = \ket{0101}, \quad \ket{b_1^3} = \ket{0011},
    \label{eq:samples}
\end{equation}
which together capture the physically relevant subspace for this Trotter step. Using these selected bitstrings, we reconstruct the ground-state energy $E_0$ and the particle number \(P\), shown in \cref{fig:mainzQC_energies_4qb} and \cref{fig:mainzQC_ps_4qb}, as functions of the background electric field \(l_0\), respectively.

As shown in \Cref{fig:mainzQC_energies_4qb}, sampling the two bitstrings in \Cref{eq:samples} is sufficient to accurately reconstruct the ground-state energy, with a relative deviation between the SKQD and exact diagonalization results at the level of \(10^{-6}\). The particle number results likewise agree with exact diagonalization, as shown in \Cref{fig:mainzQC_ps_4qb}. In particular, the particle number exhibits a jump to \(\langle P \rangle = 2\) once the background field exceeds its critical value \(l_{0,c}\), consistent with theoretical expectations (see \Cref{sec:schwinger}). Due to finite-volume effects, the discontinuous jump is smoothed, and the observed critical field \(l_{0,c}\) is larger than the continuum value of \(1/2\). A detailed discussion of the continuum limit and the extraction of \(l_{0,c}\) is given in \Cref{sec:PTlarge}, in particular \Cref{eq:l0c}.

We emphasize that the \(l_0\)-dependence of the SKQD observables shown in \Cref{fig:mainzQC_energies_4qb} and \Cref{fig:mainzQC_ps_4qb} arises solely from the classical post-processing described in steps~4 and~5 of \Cref{subsec:sample-based-krylov}. All measurements of the time-evolved state \(\ket{\psi_k}\) on the quantum processor (see \Cref{eq:psik}) are performed independently of \(l_0\), since the time-evolution operator in \Cref{eq:Ugate} contains only the kinetic part of the Hamiltonian and does not depend on the background electric field.

\begin{figure}
    \centering
    \includegraphics[width=\linewidth]{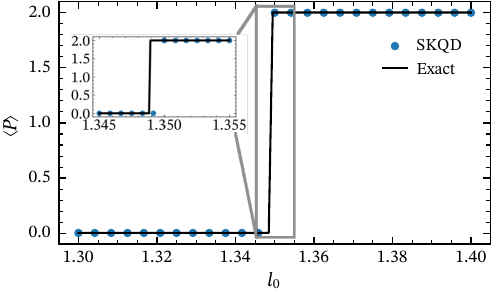}
    \caption{Particle number $\langle P\rangle$ of the Schwinger model as a function of the constant background field $l_0$, for $N=14$, $N/\sqrt{x}=30$, and $m_{\rm lat}/g=10$. Shown are SKQD results obtained on superconducting quantum processors (blue) compared with exact diagonalization (black). Results correspond to 10 Trotter steps.}
    \label{fig:IBM_N_14_ps}
\end{figure}

In summary, the proof-of-principle implementation of SKQD on the trapped-ion quantum processor with \(N=4\) qubits yields observables in excellent agreement with exact diagonalization. To investigate scalability to larger system sizes, we now turn to superconducting quantum processors.

\subsection{Superconducting Quantum Processors}
\label{subsec:IBM}

We next investigate SKQD on superconducting quantum processors from IBM to access larger system sizes, specifically \texttt{ibm\_marrakesh} and \texttt{ibm\_kingston}, which are 156-qubit devices of the Heron r2 processor type~\cite{ibm_quantum_processor_types}. We consider physical systems with $N=14$ to $20$ fermions, corresponding to the same number of qubits. While even larger systems are in principle accessible, doing so requires a careful optimization of the Trotterization parameters. Since our goal here is to demonstrate scalability while retaining a direct benchmark against exact diagonalization, we restrict the analysis to these system sizes.

\begin{figure}
    \centering
    \includegraphics[width=\linewidth]{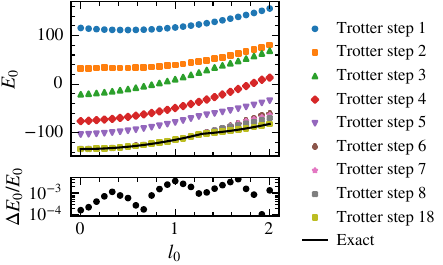}
    \caption{Ground-state energy \(E_0\) of the Schwinger model as a function of the constant background field \(l_0\), for \(N=20\), \(N/\sqrt{x}=30\), and \(m_{\rm lat}/g=10\). The upper panel shows SKQD results obtained on superconducting quantum processors for different numbers of Trotter steps (colored points) compared with exact diagonalization (black line). The lower panel displays the relative difference between the SKQD result at Trotter step 18 and the exact diagonalization result.}
    \label{fig:IBM_N_20_energy}
\end{figure}

As the initial reference state, we choose $\ket{\psi_0} = \ket{10\dots 10}$, which has minimal overlap with the true ground state. This choice is intentional, as it demonstrates that SKQD converges even when little prior physical knowledge is encoded in the initial state.

Time evolution is carried out through multiple Trotter steps~\footnote{Only data points for which the relative improvement over the previous point exceeds 1\% are retained. In principle, one may sample as many states as desired, which enlarges the Krylov subspace.}, using a time step of $dt = 0.2$ for $N = 14$ and $16$, and $dt = 0.3$ for $N = 18$ and $20$. Note, as mentioned above, other choices of $dt$ are possible. 

For the measurements, we use $n_\text{shots}=1{,}000$ for $N = 14$, $16$, and $18$, and $n_\text{shots}=10{,}000$ for $N = 20$, since the latter case requires sampling a substantially larger number of contributing bitstrings. Using the measured bitstrings from the retained Trotter steps, we construct the corresponding Krylov subspace, project the Hamiltonian into this space, and obtain the energy spectrum by classical diagonalization.

In \Cref{fig:IBM_N_14_energy}, we show the SKQD results for the ground-state energy for $N=14$ qubits across different numbers of Trotter steps. For Trotter steps 1 to 6, the SKQD results deviate significantly from the exact value. Starting from step 7, the results begin to converge, and by step 10 they agree with exact diagonalization at the $10^{-3}$ level (see \Cref{tab:energy_deviation}). Results for steps 8, 9, and $> 10$ are not shown, as the relative improvement in energy compared to the previous step falls below the threshold $c=10^{-2}$, as discussed in \Cref{sec:methods}.

In \Cref{fig:IBM_N_14_ps}, the particle number results for $N=14$ also show excellent agreement with exact diagonalization, allowing for an accurate determination of the critical background electric field $l_{0,c}$, with a relative deviation on the $10^{-4}$ level (see inset). Similar to the $N=4$ results obtained on trapped-ion quantum processors in \Cref{subsec:tiqc}, the particle number exhibits a discontinuous jump to \(\langle P \rangle = 2\) once the background field exceeds its critical value \(l_{0,c}\). Unlike the $N=4$ case, where the jump was smoothed by finite-volume effects, the jump for $N=14$ is truly discontinuous, as expected for larger system sizes. Furthermore, the observed value of \(l_{0,c}\) is smaller than for $N=4$, consistent with the expectation that finite-volume effects increase \(l_{0,c}\) (see \Cref{sec:PTlarge}).

In addition to accuracy, another key metric for SKQD is the reduction of the subspace dimension relative to the full Hilbert space. For the system sizes considered in this work, the dimension of the full Hilbert space is
\begin{equation}
    N_\text{phys} = \begin{pmatrix}
        N \\ N/2
    \end{pmatrix} = \begin{cases}
        3,\!432 \quad \text{for } N=14, \\
        12,\!870 \quad \text{for } N=16, \\
        48,\!620 \quad \text{for } N=18, \\
        184,\!756 \quad \text{for } N=20.
    \end{cases}
\end{equation}

\begin{figure}
    \centering
    \includegraphics[width=\linewidth]{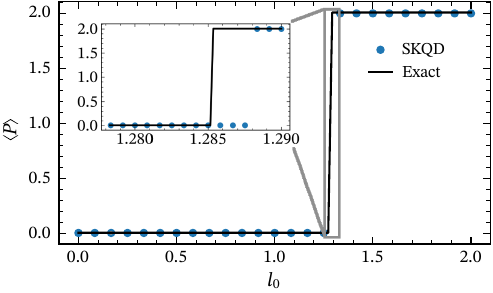}
    \caption{Particle number $\langle P\rangle$ of the Schwinger model as a function of the constant background field $l_0$, for $N=20$, $N/\sqrt{x}=30$, and $m_{\rm lat}/g=10$. Shown are SKQD results obtained on superconducting quantum processors (blue) compared with exact diagonalization (black). Results correspond to 18 Trotter steps.}
    \label{fig:IBM_N_20_ps}
\end{figure}

As discussed in \Cref{subsec:sample-based-krylov}, the Krylov space is typically much smaller than the full physically relevant Hilbert space. To examine the subspace dimension at each Trotter step, \Cref{fig:IBM_N_14_subspace} in Appendix~\ref{app:SubspaceDim} shows the number of sampled basis states as a function of the Trotter step for $N=14$ qubits. At Trotter step 10, slightly more than $1,\!000$ basis states are sampled. Since Trotter step 10 exhibits excellent agreement with the exact energy in \Cref{fig:IBM_N_14_energy}, this demonstrates that the Krylov subspace captures the relevant physics while reducing the effective Hilbert space dimension by more than 60\%.

Going to larger system sizes,  \cref{fig:IBM_N_20_energy} shows SKQD results for the ground-state energy for $N=20$. The SKQD results converge to the exact result for $k=18$, with errors at the $10^{-3}$ level  (see \Cref{tab:energy_deviation}). Note that the results for Trotter steps 9--17 and $> 18$ are not displayed, because the relative improvement in the energy compared to the previous Trotter step was smaller than $c=10^{-2}$, as discussed in \Cref{sec:methods}. 

The particle number results for $N=20$ in \cref{fig:IBM_N_20_ps} also show excellent agreement with the exact results, analogous to the $N=14$ case. The critical background electric field $l_{0,c}$ can be extracted with similarly high accuracy, although the relative deviation is slightly larger here, at the level of $10^{-3}$ (see inset).

\begin{figure}
    \centering
    \includegraphics[width=\linewidth]{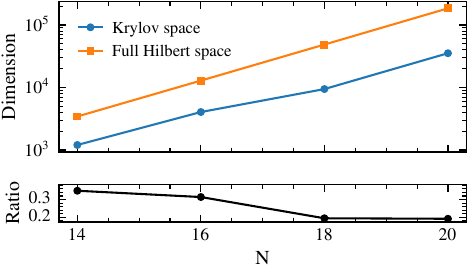}
    \caption{Krylov space dimension $\mathrm{dim}\mathcal{K}$ obtained with SKQD (blue) compared to the full Hilbert space dimension $\mathrm{dim}\mathcal{H}$ (orange) as a function of system size $N$. The lower panel shows the ratio $\mathrm{dim}\mathcal{K} / \mathrm{dim}\mathcal{H}$.}
    \label{fig:SuppressionAnalysis}
\end{figure}

Results for the reduction of the subspace dimension relative to the full Hilbert space for $N=20$ are shown in \cref{fig:IBM_N_20_subspace} in Appendix~\ref{app:SubspaceDim}. Approximately $20\%$ of the full Hilbert space is required to approximate the ground state with the precision discussed above. In \Cref{fig:SuppressionAnalysis}, we display the scaling of the sampled Krylov subspace dimension $\mathrm{dim}\mathcal{K}$ (blue) as a function of $N$, for $N$ ranging between 14 and 20. Comparing this to the scaling of the full Hilbert space dimension $\mathrm{dim}\mathcal{H}$ (orange), and in particular examining their ratio $\mathrm{dim}\mathcal{K}/\mathrm{dim}\mathcal{H}$ (lower panel), we observe that while the subspace dimension still grows exponentially, the growth is substantially suppressed relative to the full Hilbert space. The results for $N=16$ and $N=18$, used in this comparison, are provided in Appendix~\ref{app:16_18}. 

In \cref{tab:energy_deviation}, we summarize our results for different system sizes $N$: 
(i) the number of Trotter steps $k_{\mathrm{max}}$ required to reach the target error threshold $c = 10^{-2}$, which does not exhibit a clear trend as $N$ increases; 
(ii) the mean relative deviation $\overline{\Delta E_0 / E_0}$ of the SKQD ground-state energy from exact diagonalization (averaged over $l_0 \in \{0,2\}$), which remains at the $10^{-3}$ level for all $N$ considered; 
and (iii) the ratio $\mathrm{dim}\mathcal{K} / \mathrm{dim}\mathcal{H}$ between the Krylov subspace and the full Hilbert space dimensions, which decreases as $N$ increases (see \Cref{fig:SuppressionAnalysis}).

In summary, the SKQD algorithm enables a substantial reduction of the subspace required to approximate the ground state, while maintaining high accuracy with only a modest number of Trotter steps. In \Cref{sec:PTlarge}, we further present results for larger system sizes, up to $N=30$ qubits, employing a ground-state initialization of the mass Hamiltonian.

\renewcommand{\arraystretch}{1.3}
\setlength{\tabcolsep}{15.5pt}
\begin{table}[h]
\centering
\caption{Summary of results for different system sizes $N$: the number of Trotter steps $k_{\mathrm{max}}$ required to reach the error threshold $c = 10^{-2}$, the mean relative deviation $\overline{\Delta E_0 / E_0}$ of the SKQD ground-state energy from exact diagonalization (averaged over $l_0 \in \{0,2\}$), and the ratio $\mathrm{dim}\mathcal{K} / \mathrm{dim}\mathcal{H}$ between the respective Krylov and full Hilbert space dimensions.}
\begin{tabular}{c c c c}
\hline
$N$ & $k_{\rm max}$ & $\overline{\Delta E_0 / E_0}$ & $\text{dim}\mathcal{K}/\text{dim}\mathcal{H}$ \\
\hline
14 & 10 & $2.1\times 10^{-3}$ & 0.35 \\
16 & 18 & $1.7\times 10^{-3}$ & 0.31 \\
18 & 7 & $1.5 \times 10^{-3}$ & 0.19 \\
20 & 18 & $1.4\times 10^{-3}$ & 0.19 \\
\hline
\end{tabular}
\label{tab:energy_deviation}
\end{table}

\section{Results using State Initialization of the Mass Hamiltonian}
\label{sec:PTlarge}

Here, we discuss an approach for accessing larger system sizes beyond $L=20$ using a different ground-state initialization. The results in \cref{subsec:IBM} used an arbitrary initial state in the zero-charge sector, which had negligible overlap with the ground state in both phases ($l_0 < l_{0,c}$ and $l_0 > l_{0,c}$). While this demonstrates the general applicability of the algorithm, it requires sampling a large number of states to reach the correct ground state in both phases.

As an alternative, the system can be initialized in the ground state of the mass Hamiltonian, i.e., the Hamiltonian in \cref{Staggered_hamiltonian_final} with all terms set to zero except for the one proportional to $m_{\rm lat}$. This ground state, $\ket{01\dots01}$, has a large overlap with the phase at small $l_0$, though almost none with the other phase. Starting from this state drastically reduces the number of bitstrings that need to be sampled to obtain a good approximation of the ground state in both phases. This approach enables simulations of larger systems of up to $N = 30$ qubits, a regime that is challenging for standard exact diagonalization methods.

\begin{figure}
    \centering
    \includegraphics[width=\linewidth]{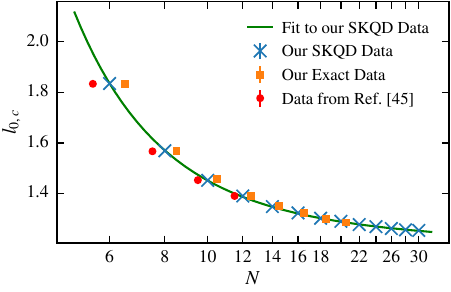}
    \caption{Critical background electric field $l_{0,c}$ as a function of the system size $N$, calculated with SKQD using the ground-state initialization of the mass Hamiltonian (blue) for $N/\sqrt{x}=30$ and $m_\text{lat}/g=10$. For comparison, VQE results from Fig.~3 of Ref.~\cite{Angelides:2023noe} are shown in red, and exact diagonalization results are shown in orange. To improve visibility, the datasets from Ref.~\cite{Angelides:2023noe} and exact diagonalization have been shifted along the x-axis by $\Delta N = \pm 0.5$, respectively. A fit to the SKQD data is shown in green. The uncertainties, smaller than the markers, reflect the spacing of $l_0$ values used to determine the transition point $l_{0,c}$.}
    \label{fig:PTlarge}
\end{figure}

To analyze the scaling of electric background field at the critical point, $l_{0,c}$, with the number of qubits $N$, we use the superconducting processor \texttt{ibm\_marrakesh} with varying numbers of shots and $dt = 0.3$, and determine the phase transition point $l_{0,c}$ from the ground-state particle number $P$ for different system sizes up to $N=30$, see \cref{fig:PTlarge}. At this size, the physical Hilbert space has dimension $N_\text{phys} = 155,\!117,\!520$. We retain only the Trotter steps for which the relative difference between consecutive steps is less than \SI{1}{\percent}.

As shown in \cref{fig:PTlarge}, our SKQD implementation with the ground-state initialization of $\ket{01\dots01}$ accurately approximates the critical background field $l_{0,c}$. For smaller systems, our results are in excellent agreement with results from quantum hardware inference runs using the VQE algorithm~\cite{Angelides:2023noe}, as well as with our exact diagonalization results. While exact solutions become impractical for $N>20$, exact diagonalization for $N \leq 20$ confirms that the phase transition point remains very close to its true value, with a relative deviation of only $\Delta l_{0,c}/l_{{0,c}}=4\times 10^{-3}$.

To compare with analytical results, we use Eq.~(11) from the supplementary material of Ref.~\cite{Angelides:2023noe}, which predicts the phase transition point as
\begin{equation}
l_{0,c} = \frac{1}{15}\left( \frac{m_\text{lat}}{g}+{\rm MS}(N) \right)\frac{1}{1-N^{-1}} + \frac{1}{2}\,,
\label{eq:l0c}
\end{equation}
for our chosen parameters. Since the mass shift (MS) in \Cref{eq:l0c} is not known analytically, we model it as
\begin{equation}
{\rm MS}(N) = \frac{a}{\sqrt{N}} + \frac{b}{N} + \frac{c}{N^2}\,.
\end{equation}
The resulting fit, shown in \cref{fig:PTlarge}, is in good agreement with the SKQD data. The fit parameters are $a = 6.5 \pm 0.2$, $b = -17 \pm 1$ and $c = 246 \pm 5$.

In \Cref{eq:l0c}, we fixed the lattice volume to $N/\sqrt{x}=30$ following Ref.~\cite{Angelides:2023noe}. Therefore, the $N\to\infty$ limit in \cref{fig:PTlarge} does not correspond to the proper continuum limit, which requires first taking $N\to\infty$ at fixed $ag$ and then $ag\to0$ (cf.\ App.\ B of Ref.~\cite{Funcke:2019zna}). In that correctly taken limit, $l_{0,c}$ would approach the continuum value $1/2$, as discussed in \cref{sec:cont}.

\begin{figure}
    \centering
    \includegraphics[width=\linewidth]{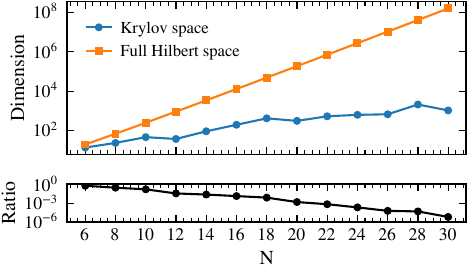}
    \caption{Krylov space dimension $\mathrm{dim}\mathcal{K}$ obtained with SKQD using the ground-state initialization of the mass Hamiltonian (blue) compared to the full Hilbert space dimension $\mathrm{dim}\mathcal{H}$ (orange) as a function of system size $N$. The lower panel shows the ratio $\mathrm{dim}\mathcal{K} / \mathrm{dim}\mathcal{H}$.}
    \label{fig:SuppPTlarge}
\end{figure}

Finally, we note that the sampled Krylov space~$\mathcal{K}$ is much smaller than the full physical Hilbert space $\mathcal{H}$, as shown in \cref{fig:SuppPTlarge}. The ratio $\mathrm{dim}\mathcal{K} / \mathrm{dim}\mathcal{H}$ of the Krylov space to full Hilbert space dimension becomes as small as $1.7 \times 10^{-3}$ for $N=20$ and $6.8 \times 10^{-6}$ for $N=30$. This further demonstrates that SKQD can substantially reduce the Hilbert space. Although the Krylov space dimension still scales exponentially, its significantly slower increase underscores the potential of the method for simulating lattice gauge theories in larger volumes.

\section{Conclusion}
\label{sec:conclusion}

In this work, we presented the first application of Sample-Based Krylov Quantum Diagonalization (SKQD) to lattice gauge theories executed on current quantum hardware. Using the Schwinger model with a $\theta$-term as a benchmark, we applied SKQD to analyze the phase diagram and extract ground-state energies and particle numbers, enabling a precise determination of the critical background electric field $l_{0,c}$. We implemented SKQD on both trapped-ion and superconducting quantum processors for system sizes ranging from $N=4$ to $N=30$ qubits, achieving a substantial reduction of the effective subspace needed to accurately approximate the ground state. Across all system sizes considered, our results show excellent agreement with exact diagonalization. 

For smaller system sizes, $N=4$ to $N=20$ qubits, we employed an arbitrary initial state and found that SKQD reduced the number of basis states by up to $\sim 80\%$ (see \cref{tab:energy_deviation}). Furthermore, by initializing the algorithm in the ground state of the mass Hamiltonian, we extended the simulations to larger system sizes of up to $N=30$ qubits. This initialization enabled an accurate determination of the critical background field $l_{0,c}$, while reducing the number of sampled basis states by several orders of magnitude relative to the full Hilbert space (down to $6.8 \times 10^{-6}$; see \cref{fig:SuppPTlarge}). 

Although the Krylov space dimension still increases exponentially with system size, the significantly slower growth compared to the full Hilbert space highlights the potential of SKQD for scaling to larger quantum systems. The method shows enhanced robustness to noise compared with alternative approaches such as KQD or VQE, making it particularly attractive in the NISQ area. Furthermore, as we have shown, physical symmetries can be exploited in classical post-processing, and the quantum resources required can be substantially reduced by restricting time evolution to the kinetic Hamiltonian rather than the full Hamiltonian.

Overall, our results demonstrate that SKQD is a suitable and resource-efficient method for studying lattice gauge theories on current quantum devices. Future work will focus on extending the method to larger system sizes, excited states, as well as higher-dimensional lattice gauge theories.

\section*{Contributions}
ER proposed the project and conducted the theoretical background study. ER and JE refined and adapted existing theoretical and methodological frameworks for the specific problem considered. ER performed the analysis on superconducting quantum processors, and JE carried out the trapped-ion analysis. 
ER and JE drafted the manuscript. LF contributed to writing, proofreading, and structural editing.
LF, FSK, and UP supervised the project and contributed to the manuscript’s revision and finalization.

\begin{acknowledgments}
We wish to thank Simran Singh for valuable discussions.
The authors gratefully
acknowledge the granted access to the Marvin cluster
hosted by the University of Bonn.
This project was funded by the Deutsche Forschungsgemeinschaft (DFG, German Research Foundation) as part of the CRC 1639 NuMeriQS -- project no.\ 511713970 and under Germany's Excellence Strategy – Cluster of Excellence Matter and Light for Quantum Computing (ML4Q) EXC 2004/1 – 390534769.
The JGU team acknowledges financial support by the BMBF within the projects SYNQ, IQuAn and ATIQ. 
This work is part of the Quantum Computing for High-Energy Physics (QC4HEP) working group.
We acknowledge the use of IBM Quantum services for this work. The views expressed are those of the authors, and do not reflect the official policy or position of IBM or the IBM Quantum team.
\end{acknowledgments}

\section*{Data Availability}

The data that support the findings of this article are
openly available \cite{OurData}.

\FloatBarrier

\bibliography{skqd}
\FloatBarrier
\newpage

\appendix

\section{Experimental Details of Implementation on Trapped-Ion Quantum Processor}
\label{sec:QC}
In this appendix, we provide details on the experimental implementation of the SKQD algorithm on a shuttling-based trapped-ion quantum computer \cite{kaufmann2017scalable}, based on a linearly segmented Paul trap \cite{Kielpinski:2002wbd,Kaushal:2019skm,Moses:2023ozv}. 
Trapped-ion quantum computers offer several beneficial features, including exceptionally long coherence times and enhanced qubit connectivity. Furthermore, ions are inherently identical, enabling stable and reproducible control.
Architecture, software and characteristics of the platform employed for this work are briefly discussed in this section.

To realize qubits using the electronic structure of $^{40}$Ca$^+$ ions, we choose the two Zeeman levels of the $S_{1/2}$-groundstate as the two qubit states, differing only in the orientation of the valence electron spin, with the assignment \[\ket{S_{1/2}, m_j=+1/2} \rightarrow \ket{0}, \quad  \ket{S_{1/2}, m_j=-1/2} \rightarrow \ket{1}\] and a splitting of around $2 \pi \times $\SI{10}{\mega \hertz}. 
Optically driven transitions of the Calcium ions are used for cooling, state preparation, measurements, as well as single- and two-qubit gates.

In the trap, the ions are stored in Coulomb-crystals with a maximum size of two ions per crystal. 
The storage configuration of the ions within the trap can be dynamically changed via \textit{split-, merge-, swap-} and \textit{shuttle-}operations. 
One segment, the \textit{Laser Interaction Zone} (LIZ) is designated to the optical fields interacting with the ions. The LIZ is where cooling, splitting, merging, swapping, as well as all other gate operations - including state preparation and readout - take place.

One-qubit gates are realised using \textit{stimulated Raman transitions}. This works via two Raman beams, off-resonant from the $\ket{S_{1/2}} \leftrightarrow \ket{P_{1/2}}$ transition by about \SI{800}{\giga \hertz}. This allows the realisation of the single-qubit rotations: 

\begin{equation}
\label{eq:one-qb-gate}
    R(\theta, \phi) = \exp\Big\{-\frac{i\theta}{2} (\cos \phi \, X + \sin \phi \, Y)\Big\},
\end{equation}
where the angles $\theta$ and $\phi$ are set via the variation of the laserpower and relative phase between the two Raman-beams, and $X, Y$ are Pauli matrices. The z-rotations on our hardware are kept purely virtual. This means that the z gates - changing only the relative phase between the two basis states $\ket{0}, \ket{1}$, without any population transfer - is accounted for by keeping track of the phase for all ions, and shifting phases of rotations accordingly.
We use a constant magnetic field of about \SI{0.37}{\milli \tesla} to realize the Zeeman splitting of the $S_{1/2}$-groundstate. 
The magnetic field is not constant along the trap axis. 
Since magnetic field inhomogeneities in conjunction with shuttling operations lead to the accumulation of spurious phases, we correct phases based on the measurement of the magnetic field along the trap axis. 

The two-qubit gate operates by coupling the ions' internal electronic states to their collective motional modes, generating entanglement through controlled motion. The unitary of the native two-qubit gate, acting on two qubits $q_m$, $q_n$ is given by: 
\begin{equation}
    G_{mn}(\theta) = \exp \Big\{ -i \frac{\theta}{2} Z_m \otimes Z_n \Big\},
\end{equation}
where $Z_n, Z_m$ are Pauli-$Z$ gates acting on qubit $q_n$, and $q_m$, respectively. 
To ensure that the desired phase is accumulated accurately during the gate operation, the ions must be cooled well below the Doppler limit of approximately \SI{550}{\micro\kelvin} (corresponding to about 7 phonons) \cite{Hilder2022}. 
This is achieved using motional sideband transitions. 
In particular, \textit{sideband cooling} (SB) allows the ions to be cooled from the Doppler limit down to the motional ground state, minimizing thermal excitations that would otherwise degrade gate fidelity. 
Typical gate fidelities for the two-qubit gate are \SI{99.6(2)}{\percent} \cite{Hilder:2021wde}.

Prior to measurement, the entire quantum computing system—including laser beams, control electronics, and trapping potentials—must be carefully calibrated to compensate for slow drifts in laser frequency, intensity, and magnetic field. 

The measurement workflow begins with submitting the quantum circuit in OpenQASM 3 format \cite{Cross:2021bcz}, which is transferred to the quantum hardware via a multi-layer software stack.
At the initial stage, the circuit is processed by the circuit compiler \cite{Kreppel:2022oyr}, where it undergoes validation to confirm that it conforms to the operational constraints of the target device. After validation, the compilation proceeds in two principal phases.
First, the circuit is transpiled into the native gate set supported by the hardware. During this process, phase-tracking routines and a collection of \verb|pytket| \cite{pytket} compilation passes are employed to optimize the circuit at the gate level. 

After mapping the circuit onto the physical architecture, a corresponding shuttling sequence is generated and optimized using a dedicated shuttling compiler \cite{Durandau:2022rva}, which minimizes ion shuttling overhead and thus reduces the overall runtime of the algorithm. The finalized low-level instructions are subsequently executed by a real-time hardware control system.
After execution, the high-level control server securely retrieves the experimental outcomes, and the collected measurement data is archived in a relational database.

\begin{figure}
    \centering
    \includegraphics[width=\linewidth]{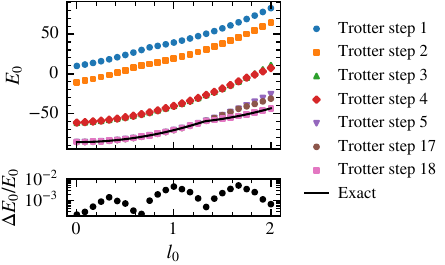}
    \caption{Ground-state energy \(E_0\) of the Schwinger model as a function of the constant background field \(l_0\), for \(N=16\), \(N/\sqrt{x}=30\), and \(m_{\rm lat}/g=10\). The upper panel shows SKQD results obtained on superconducting quantum processors for different numbers of Trotter steps (colored points) compared with exact diagonalization (black line). The lower panel displays the relative difference between the SKQD result at Trotter step 18 and the exact diagonalization result.}
    \label{fig:GS_16}
\end{figure}

\begin{figure}
    \centering
    \includegraphics[width=\linewidth]{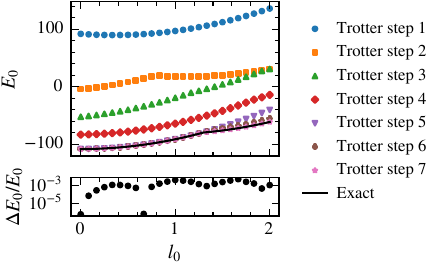}
    \caption{Ground-state energy \(E_0\) of the Schwinger model as a function of the constant background field \(l_0\), for \(N=18\), \(N/\sqrt{x}=30\), and \(m_{\rm lat}/g=10\). The upper panel shows SKQD results obtained on superconducting quantum processors for different numbers of Trotter steps (colored points) compared with exact diagonalization (black line). The lower panel displays the relative difference between the SKQD result at Trotter step 7 and the exact diagonalization result.}
    \label{fig:GS_18}
\end{figure}

\begin{figure}
    \centering
    \includegraphics[width=\linewidth]{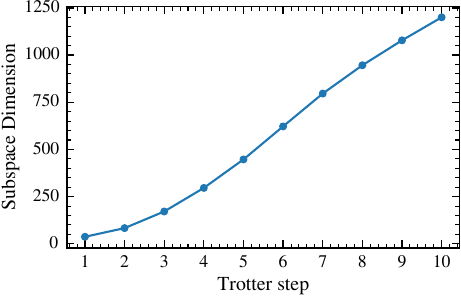}
    \caption{Size of the subspace of the sampled basis states against the number of Trotter steps for $N=14$ qubits. In this case the physical Hilbert space has $3,\!432$ basis states.}
    \label{fig:IBM_N_14_subspace}
\end{figure}

\begin{figure}
    \centering
    \includegraphics[width=\linewidth]{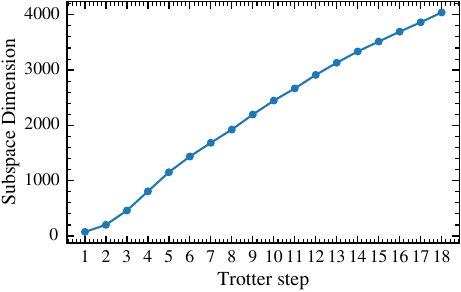}
    \caption{Size of the subspace of the sampled basis states against the number of Trotter steps for $N=16$ qubits. In this case the physical Hilbert space has $12,\!870$ basis states.}
    \label{fig:IBM_N_16_subspace}
\end{figure}

\begin{figure}
    \centering
    \includegraphics[width=\linewidth]{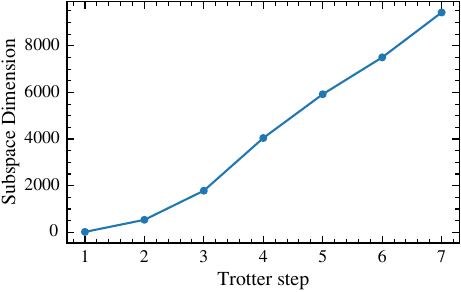}
    \caption{Size of the subspace of the sampled basis states against the number of Trotter steps for $N=18$ qubits. In this case the physical Hilbert space has $48,\!620$ basis states.}
    \label{fig:IBM_N_18_subspace}
\end{figure}

\begin{figure}
    \centering
    \includegraphics[width=\linewidth]{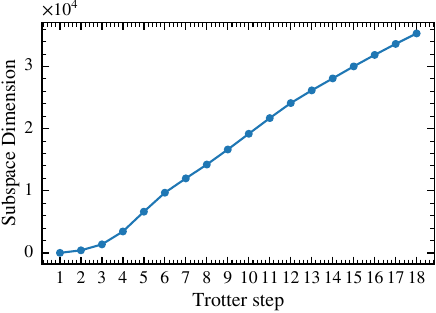}
    \caption{Size of the subspace of the sampled basis states against the number of Trotter steps for $N=20$ qubits. In this case the physical Hilbert space has $184,\!756$ basis states.}
    \label{fig:IBM_N_20_subspace}
\end{figure}

\section{Additional Results for the Ground-State Energy for $N=16$ and $N=18$ Qubits}
\label{app:16_18}

Here, we provide additional experimental results obtained on the \texttt{ibm\_marrakesh} and \texttt{ibm\_kingston} quantum processors, complementing the data provided in \Cref{subsec:IBM}. While the main text reported the ground-state energy approximations for systems with $N = 14$ and $N = 20$ qubits, \cref{fig:GS_16,fig:GS_18} shows corresponding results for $N = 16$ and $N = 18$. These data points are included to complete the scaling analysis shown in \cref{fig:SuppressionAnalysis}.

\section{Scaling of the Subspace Dimension with the Number of Trotter Steps}
\label{app:SubspaceDim}

In this appendix, we present additional plots illustrating the relationship between the number of Trotter steps (for the experimental results shown in \cref{subsec:IBM} and \cref{app:16_18}) and the dimension of the subspace spanned by the sampled Krylov basis states. In \cref{subsec:IBM} and \cref{app:16_18}, we demonstrated how the ground-state energy approximation converges to the exact value as the number of Trotter steps increases. However, since a single Trotter step does not correspond to a fixed number of sampled states, we explicitly display the subspace dimension reached after each Trotter step in \cref{fig:IBM_N_14_subspace,fig:IBM_N_16_subspace,fig:IBM_N_18_subspace,fig:IBM_N_20_subspace} for $N = 14$, $N = 16$, $N = 18$, and $N = 20$.

\end{document}